\documentclass[twocolumn,showpacs,amsmath,amssymb,aps,floatfix]{revtex4}

\usepackage{graphicx}
\usepackage{dcolumn}
\usepackage{bm}

\begin{document}
\title{The role of conditional probability in multi-scale stationary Markovian processes}        
\author{Salvatore Miccich\`e}
\affiliation{Dipartimento di Fisica e Tecnologie Relative, Universit\`a degli Studi di Palermo, \\
             Viale delle Scienze, Ed. 18, I-90128 Palermo, Italy}

\date{\today}

\begin{abstract} 

The aim of the paper is to  understand how the inclusion of more and more time-scales into a stochastic stationary Markovian process affects its conditional probability. To this end, we consider two Gaussian processes: $(i)$ a short-range correlated process with an infinite set of time-scales bounded from below, and $(ii)$ a power-law correlated process with an infinite and unbounded set of time-scales. For these processes we investigate the equal position conditional probability $P(x,t|x,0)$ and the mean First Passage Time $T_x(\Lambda)$. The function $P(x,t|x,0)$ can be considered as a proxy of the persistence, i.e. the fact that when a process reaches a position $x$ then it spends some time around that position value. The mean First Passage Time  can be considered as a proxy of how fast is the process in reaching a position at distance $\Lambda$ starting from position $x$.

In the first investigation we show that the more time-scales the process includes, the larger the persistence. Specifically, we show that the power-law correlated process shows a slow power-law decay of $P(x,t|x,0)$ to the stationary pdf. By contrast, the short range correlated process shows a decay dominated by an exponential cut-off. Moreover, we also show that the existence of an infinite and unbouded set of time-scales is a necessary and not sufficient condition for observing a slow power-law decay of $P(x,t|x,0)$. In fact, in the context of stationary Markovian processes such form of persistence seems to be associated with the existence of an algebraic decay of the autocorrelation function. In the second investigation, we show that for large values of $\Lambda$ the more time-scales the process includes, the larger the mean First Passage Time, i.e. the slowest the process. On the other hand, for small values of $\Lambda$, the more time-scales the process includes, the smaller the mean First Passage Time, i.e. when a process statistically spends more time in a given position the likelihood that  it reached nearby positions by chance is also enhanced.


%
\end{abstract}

\maketitle
%
%
%
\section{Introduction}   \label{intro}

Stochastic processes are used to model a great variety of systems in disciplines as disparate as physics \cite{BM,VanKampen81,risken,gardiner,Schuss,Oksendal}, 
genomics \cite{Waterman,Durbin}, finance \cite{Bouchaud,Mantegna}, climatology \cite{vanStorch} and social sciences \cite{Helbing}. 

Among others, Markov processes play a central role in the modeling of 
natural phenomena due to their simplicity. In fact, a Markov process is fully determined 
by the knowledge of its probability density function (pdf)
$W(x,t)$  and its conditional transition probability
$P(x_{n+1},t_{n+1}|x_{n},t_{n})$. When a stationary Markovian process is continuous both in space and time, the time evolution of these distributions is described by a Fokker-Planck (FP)
equation \cite{risken,VanKampen81}. In this paper we will only consider stationary Markovian processes that can be described by using a FP equation. 

Stochastic processes can be considered from the point of view of their correlation
properties. Under this respect,
random variables are divided in short-range and 
long-range correlated variables. Short-range 
correlated variables are characterized by a finite mean of time-scales 
of the process whereas a similar mean time-scale 
does not exist for long-range correlated variables \cite{Beran94, anomalous1, anomalous2,Oliveira,Chavanis}. 

Several stationary Markovian processes are short-range correlated. In fact, the paradigmatic stationary Markovian process is the Ornstein--Uhlenbeck (OU) one \cite{ornstein}, whose autocorrelation function is the exponential function $\rho(\tau) = e^{- \tau/T}$ where $T$ is the time-scale of the process. However, generally speaking a stationary Markovian process can be multi-scale, i.e. it may admit either a discrete or a continuum set of time-scales \cite{GS1,SK,GS2,HL,Zoller}. In general, the autocovariance function $R(\tau)=\langle x(t) x(t+\tau) \rangle$  of an additive stationary Markovian process admitting a Fokker--Planck equation can be written as  $R(\tau)=\sum_{n=0}^{\lambda_p}  c_n^2 \exp(- \lambda \tau)+\int_{\lambda_c}^\infty c_\lambda^2 \exp(- \lambda \tau)$, where ${\lambda_0=0,\lambda_1,...,\lambda_p}$ and $]\lambda_c,+\infty[$  ($\lambda_c \geqslant \lambda_p$)  constitute the discrete and continuum parts of the eigenvalue spectrum of the FP equation. The quantities $c_n$, $c_\lambda$ are appropriate weights computed in terms of the eigenfunctions \cite{risken}. This picture indicates that the more time-scales included in the process, the slower the decay of the autocorrelation function. A generalization to non additive processes that are obtained by performing appropriate coordinate transformations of the additive ones has been considered in Ref. \cite{paperpa19}.

Our main interest is to investigate what are the determinants of a long-range correlated stationary Markovian process, at the level of its probability distributions. In Ref. \cite{paperpa19} we considered stationary Markovian stochastic processes with a power-law decaying autocorrelation function and a stationary pdf with Gaussian or exponential tails. Those results suggested that the stationary pdf does not bear any special feature with respect to the  correlation properties of the process. We therefore want to investigate how the inclusion of more and more time-scales into stationary Markovian stochastic processes affects their conditional probability, which is the other probability distribution that fully characterizes the process. 

To this end we will consider two stationary Markovian processes with Gaussian pdf and autocorrelation function not described by a single exponential. In one case the process is short--range correlated and in the other case the autocorrelation function decays like a power-law. We preliminarily show that, as suggested by the Doob theorem \cite{Doob}, this is only possible if one admits a conditional probability different from Gaussian for both processes. We then investigate the conditional probability by following two complementary approaches. First we consider the conditional probability at equal spatial positions $P(x,t|x,0)$ as a proxy of how persistent is a process in a given position. We show that the higher the number of time-scales incorporated in the stochastic process the higher the persistence. 

We then consider a quantity related to the conditional probability, i.e. the mean First Passage Time $T_x(\Lambda)$ \cite{gardiner}. It can be considered as a proxy of how fast is the process in reaching a position at distance $\Lambda$ from the initial position $x$. We show that the higher the number of time-scales incorporated in the stochastic process the slower the process is.  On the other hand, for small values of $\Lambda$, the more time-scales the process includes, the smaller the mean First Passage Time, i.e. when a process statistically spends more time in a given position the likelihood that  it reached nearby positions by chance is also enhanced.

The paper is organized as follows: in section \ref{gauss} we will show two examples of processes that are Gaussian distributed and whose autocorrelation function is not exponential. In one case, see section \ref{riskgaussall}, the process is short-range correlated. In the second case, see section \ref{chigaussall}, the process can be long-range correlated with an appropriate choice of the relevant parameters. In any case the process is always power-law correlated. In section \ref{gausspersistence} we consider the conditional probability at equal spatial position $P(x,t|x,0)$ and show that it decays like a power-law decays for the algebraically correlated process. We also show that such peculiar behaviour is not shared by other processes that include an infinite and unbounded set of time-scales. In section \ref{gausspropagator} we consider the mean First Passage Time $T_x(L)$ and show that the higher the number of time-scales incorporated in the stochastic process the slower the process is. Finally, in section \ref{concl} we draw our conclusions.

\section{Processes with Gaussian stationary pdf}   \label{gauss}

\subsection{A short-range correlated process}   \label{riskgaussall}

Let us consider the stochastic process described by the following Langevin equation \cite{risken}:
\begin{eqnarray}
               && \dot{x}(t)=- h(x(t))\, + D\,\Gamma(t)  \nonumber \\
               &&  h(x)=\left \{ \begin{array}{cc}
                                +k   &{\rm{if}}~~x < 0 ,\\
                                     &   \\
                                -k   &{\rm{if}}~~x >  0 .   
                        \end{array} \right. \label{D1risk}
\end{eqnarray}
where $k$ is a real constant and $\Gamma(t)$ is a $\delta$--correlated Gaussian noise term. The diffusion coefficient $D$ will be set to unity hereafter. The stationary distribution of this process is:
\begin{eqnarray}
               W_s(x)={k\over 2}\,{\rm{exp}}( -k |x|) \label{pdfrisken}
\end{eqnarray}
By using the methodology of eigenfunction expansion \cite{risken,gardiner} it is possible to prove that the autocovariance function $R(t)=\langle x(t)x(0)\rangle$ of the above process is:
\begin{eqnarray}
          &&   R_s(t)={2 \over k^2} (1 - 2 \tau+4 \tau^2+ {8 \over 3} \tau^3)\,(1-{\rm{Erf}}(\sqrt{\tau})+  \label{ACrisken} \\
          &&   \quad -\,{{4 \sqrt{\tau}}\over{3 k^2 \sqrt{\pi}}}\,(2 \tau-1)(3 + 2 \tau)\,{\rm{exp}}(- \tau)                                 \qquad 
               \tau={k^2 \over 4} t \nonumber
\end{eqnarray}
which behaves like a power-law with an exponential truncation for large time lags: $R_s(t) \approx {\rm{exp}}(-{k^2 \over 4} t) t^{-3/2}$ as $t \to \infty$.

Let us now consider the coordinate transformation:
\begin{eqnarray}
                 x \mapsto y=f_s(x)=\sqrt{2\,s}\,{\rm{Erf}}^{-1}\bigl[1-e^{- k |x|}\bigr]  \label{risken_to_gauss}
\end{eqnarray}
By using the Ito lemma, one can show that, starting from the process of Eq. $(\ref{D1risk})$, in the coordinate space $y=f_s(x)$ one gets a multiplicative stochastic process whose stationary pdf is given by:
\begin{eqnarray}
               W_g(y)={1\over{\sqrt{2 \pi s}}} {\rm{exp}}({ 1 \over{2 s}} y^2)  \label{pdfOU}
\end{eqnarray}
with $s$ an additional arbitrary parameter. The corresponding drift and diffusion coefficients are given by:
\begin{eqnarray}
                       &&       H(y)=  {1 \over 2}\,k^2\,e^{y^2/(2 s)}\,\, \Bigl(1-{\rm{Erf}}\Bigl(|y|/\sqrt{2 s}\Bigl) \Bigl)\,\,\times \label{hRGA}  \\
                       &&       \hspace{1 cm}            \times\,\Bigl(
                                                                     \pi\,e^{y^2/(2 s)}\,\Bigl(1-{\rm{Erf}}\Bigl(|y|/\sqrt{2 s}\Bigl) \Bigl) - \epsilon\,2 \sqrt{2 \pi s}
                                                             \Bigl)   \nonumber \\
                       &&       G(y)=  \sqrt{\pi \over 2}\,k\,\sqrt{s}\,e^{y^2/(2 s)}\, \Bigl(1-{\rm{Erf}}\Bigl(|y|/\sqrt{2 s}\Bigl) \Bigl)\label{gRGA} 
\end{eqnarray}
where $\epsilon=+1$ when $y>0$ and $\epsilon=-1$ when $y<0$. For large values of $y$ the drift and diffusion coefficient behave as:
\begin{eqnarray}
              H(y) \propto - k^2 s\,{1 \over y} 
                           \qquad 
              G(y) \propto + k\,s\,{1 \over y} 
                           \qquad 
              y \to + \infty. \nonumber 
\end{eqnarray}

\begin{figure} 
\begin{center}
              \resizebox{1\columnwidth}{!}{\includegraphics[scale=0.30]{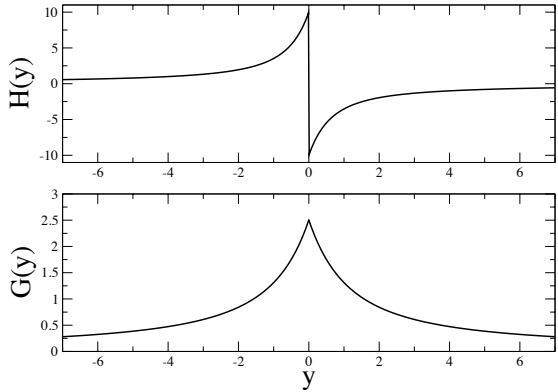} }
              \caption{The figure shows the drift coefficient $H(y)$ (top panel) and diffusion coefficient $G(y)$ (bottom panel) of the process defined by Eq. $(\ref{D1risk})$ with the coordinate transformation of Eq. $(\ref{risken_to_gauss})$ for the case when $k=2.0$ and $s=1.0$, which corresponds to a variance $\langle y(t)^2\rangle=0.5$.}  \label{DriftDiffRiskGauss}
\end{center}   
\end{figure}

In Fig. $\ref{DriftDiffRiskGauss}$ we show the drift coefficient $H(y)$ (top panel) and diffusion coefficient $G(y)$ (bottom panel) for the case when $k=2.0$ and $s=1.0$, which corresponds to a variance $\langle x(t)^2\rangle=0.5$ and $\langle y(t)^2\rangle=1.0$. The diffusion coefficient $G(y)$ is continuous in $y=0$ although its first derivative is discontinuous. The drift coefficient $H(y)$ suffers a discontinuity in $y=0$.  

According to the eigenfunction expansion methodology \cite{risken,gardiner}, the autocorrelation function of the process defined by Eq. $(\ref{D1risk})$ with the coordinate transformation of Eq. $(\ref{risken_to_gauss})$ is given by $\rho_s(\tau)= (\langle y(t+\tau)y(t)\rangle-\langle y(t)\rangle^2)/(\langle y^2(t)\rangle-\langle y(t)\rangle^2)$ where:
\begin{eqnarray}
       &&   \langle y(t) y(t+\tau) \rangle=\int_0^\infty d \lambda\,{\cal{C}}_\lambda^{2} e^{- \lambda \tau}  \label{Rrisk}  \\
       &&   {\cal{C}}_\lambda=\int_{-\infty}^{+\infty} dx\,f_s(x)\,\psi_0(x)\,\psi_\lambda(x) \nonumber
\end{eqnarray}
where $\psi_0(x)$ and $\psi_\lambda(x)$ are the eigenfunctions of the Schr\"odinger equation with potential $V_S(x)$$=k^2/4 - k\,\delta(x)$ associated to the stochastic process of Eq. $(\ref{D1risk})$. Eq. $(\ref{Rrisk})$ can be used to numerically obtain the autocorrelation of the new process defined by Eq. $(\ref{D1risk})$ with the coordinate transformation of Eq. $(\ref{risken_to_gauss})$. 

\begin{figure} 
\begin{center}
              \resizebox{1\columnwidth}{!}{\includegraphics[scale=0.30] {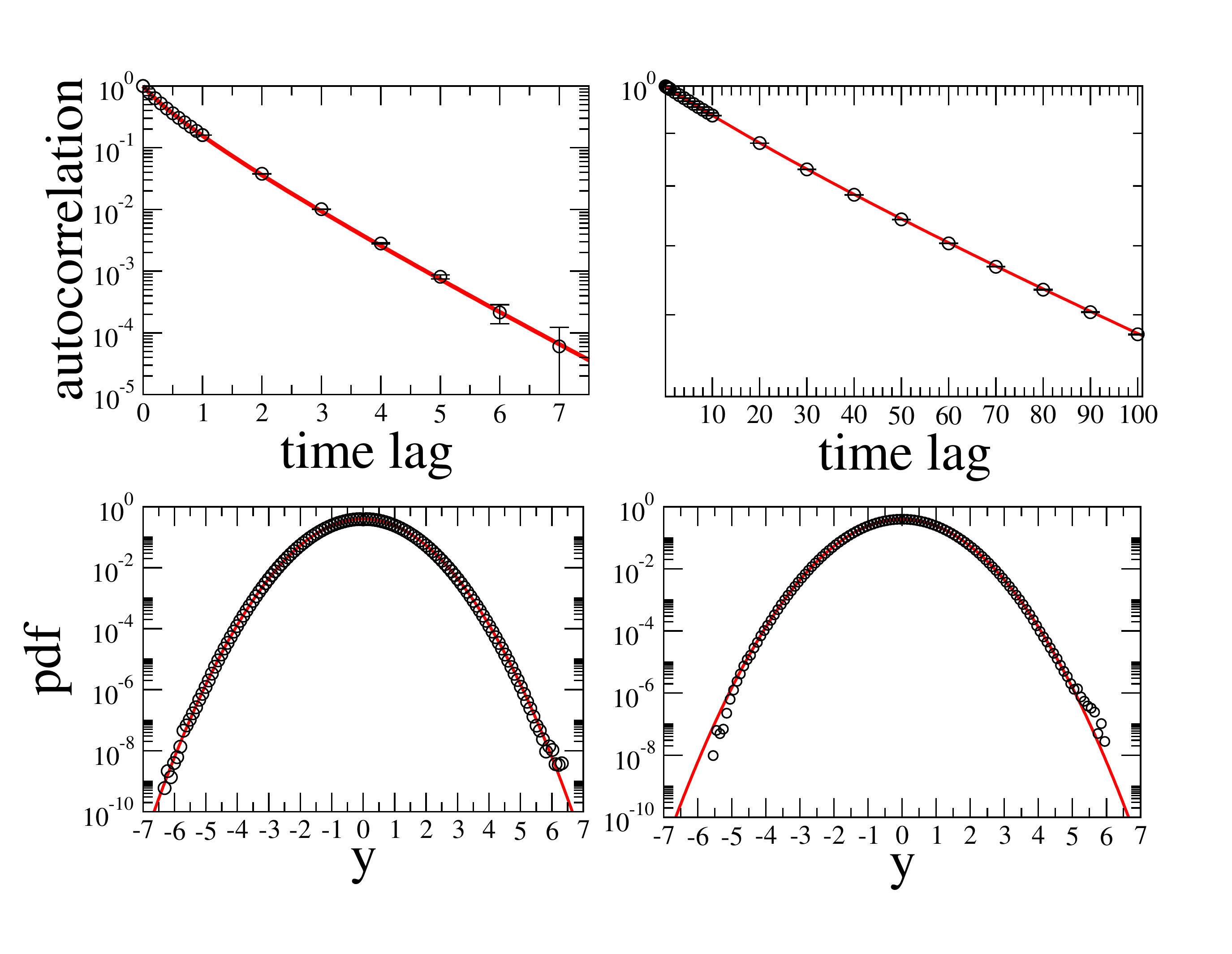} }
              \caption{(Color online) The figure shows time--average numerical simulations performed according to Eq. $(\ref{ACtimeens})$ for the case when when $s=1.0$, and $k=2.0$ (left), $k=0.1$ (right). In the top panel we show the results for the autocorrelation function. The solid (red) line shows the theoretical predictions obtained from Eq. $(\ref{Rrisk})$, while the open circles show the result of the numerical simulations. The bottom panel shows the stationary pdf of the process. Again the solid (red) line shows the theoretical prediction of Eq. $(\ref{pdfOU})$, while the open circles show the result of the numerical simulations. The parameters of the numerical simulations are given in the text.}  \label{fig:riskengauss}
\end{center}   
\end{figure}

In the top panels of Fig. $\ref{fig:riskengauss}$ we show the results of numerical simulations of the autocorrelation function performed for the case when $s=1.0$, and $k=2.0$ (left), $k=0.1$ (right). The solid (red) line shows the theoretical predictions obtained from Eq. $(\ref{Rrisk})$, while the open circles show the results of the numerical simulations. The bottom panels of Fig. $\ref{fig:riskengauss}$ show the stationary pdf of the process when $s=1.0$, and $k=2.0$ (left), $k=0.1$ (right). Again the solid (red) line shows the theoretical prediction of Eq. $(\ref{pdfOU})$, while the open circles show the result of the numerical simulations.

Those shown in Fig. $\ref{fig:riskengauss}$ are time--average numerical simulations performed according to the relation:
\begin{eqnarray}
              \rho(\tau)= {1 \over T} \int_0^T dt~x_*(t) x_*(t+\tau) \label{ACtime}
\end{eqnarray}
where $T$ is the length of the simulated time-series and $x_*(t)$ is one realization of the process. Indeed, in order to improve the statistical reliability of our numerical simulations, in the region $\tau\geq1$ we have also averaged over a number M of different realizations of the process:
\begin{eqnarray}
              \rho_T(\tau)= {1 \over M} \sum_{j=1}^M {1 \over T} \int_0^T dt~x_j(t)~x_j(t+\tau) \label{ACtimeens}
\end{eqnarray}
The data shown in the figure are the mean and the standard deviations of the $M$ autocorrelation values computed in each iteration for each time lag. The parameters of the numerical simulations of the autocorrelation function are: $M=1$ in the region $\tau \leq 1$, $M=100$ in the region $\tau \in [1,10]$  and $M=5$ in the region $\tau \geq 10$ (not shown in the panel) when $s=1.0$ and $k=2.0$. The size of each time-series was: $T=10^{9}$ with a time-step of $\Delta t=0.001$ when $\tau<10$ and $T=10^{11}$ with a time-step was $\Delta t=0.01$ when $\tau>10$. When $s=1.0$ and $k=0.1$ we used $M=1$ in the region $\tau \leq 1$, $M=10$ in the region $\tau \in [1,10]$ and $M=5$ in the region $\tau \geq 10$. The size of each time-series was: $T=10^{9}$ with a time-step of $\Delta t=0.005$ when $\tau<10$ and $T=10^{11}$ with a time-step was $\Delta t=0.01$ when $\tau>10$. The starting points of the simulated time-series were all the same with $x_j(0)=0.1$ where $j=1, \cdots, M$. The numerical simulations of the pdfs in the two bottom panels were obtained by using the same numerical data considered in the simulation of the autocorrelation function when $\tau>10$. In order to simulate the process in the $y$ coordinate space, we started by simulating the process of Eq. $(\ref{D1risk})$ and computed $y=f_s(x)$ for each simulated $x$ value.

By inspecting the two top panels of Fig. $\ref{fig:riskengauss}$ it should be evident that the autocorrelation of the process defined by Eq. $(\ref{D1risk})$ with the coordinate transformation of Eq. $(\ref{risken_to_gauss})$ can not be described by a single exponential, as in the case of the OU process. Indeed, by performing appropriate coordinate transformations one gets:
\begin{eqnarray}
       &&   {\cal{C}}_\lambda=2\,\sqrt{{k\,s}\over{2\,\pi}}\,\Bigl( \lambda - {k^2 \over 4} \Bigl)^{-3/4} \int_{0}^{+\infty} dy\,g(y T)\, \sin(y)  \\
       &&  g(z)=e^{- k z/2}\,{\rm{Erf}}^{-1}(1-e^{- k z})  \quad T=\Bigl( \lambda - {k^2 \over 4} \Bigl)^{-1/2} \nonumber
\end{eqnarray}
By using {\em{Lemma (6.1)}} in chapter 9 of Ref. \cite{Olver74} one can show that:
\begin{eqnarray}                         
                \int_{0}^{+\infty} dy\,g(y T)\, \sin(y) \approx{1 \over {T^2}} \qquad T \to \infty
\end{eqnarray}
Therefore, for small energy values $\lambda$ we get:
\begin{eqnarray}
       &&   {\cal{C}}_\lambda \approx \Bigl( \lambda - {k^2 \over 4} \Bigl)^{1/4} \qquad \lambda \to 0
\end{eqnarray}
This implies that for large time lags $R_s(t) \approx {\rm{exp}}(-{k^2 \over 4} t) t^{-3/2}$ as $t \to \infty$. We thus have a short-range correlated stochastic process whose autocorrelation function is not exponential.

\subsection{A power-law correlated process}   \label{chigaussall}

Let us consider the stochastic process described by the following nonlinear Langevin equation \cite{Zoller,paperpa19}:
\begin{eqnarray}
          &&  \dot{x}=h(x(t))+ \Gamma(t) \nonumber \\ 
          &&  h(x)=\left \{ \begin{array}{cc}
               -2 \sqrt{V_0} \tan (\sqrt{V_0}x) &{\rm{if}}~~|x| \leq L ,\\
                                         & \\
                 (1-\sqrt{1+4~V_1})/ x   &{\rm{if}}~~|x| >   L .   
                                  \end{array} \right. \label{D1chimera} \\
          &&  V_1=L\,\sqrt{V_0}\,\tan\bigl( \sqrt{V_0} L\bigl) \Bigl(1+L\,\sqrt{V_0}\,\tan\bigl( \sqrt{V_0} L\bigl)\Bigl)   \nonumber                 
\end{eqnarray}
where $L$ and $V_0$ are real arbitrary constants and $\Gamma(t)$ is a $\delta$--correlated Gaussian noise term. It is straightforward to show that the stationary distribution of this process is:
\begin{eqnarray}
            &&  W_l(x)=\left \{ \begin{array}{cc}
                                   N_{II} \cos(\sqrt{V_0}x) &{\rm{if}}~~|x| \leq L ,\\
                                                                             &   \\
                                   N_I 1/x^\alpha   &{\rm{if}}~~|x| >   L .   
                              \end{array} \right. \label{pdfCHI} \\
            && \alpha=\sqrt{1+4\,V_1}-1 = 2 L\,\sqrt{V_0}\,\tan\bigl( \sqrt{V_0} L\bigl)     \nonumber                
\end{eqnarray}
where $N_I$ and $N_{II}$ are real constants that can be obtained by imposing that $W_l(x)$ is continuous and normalized to unity. By using the methodology of eigenfunction expansion \cite{risken,gardiner} it is possible to prove that the autocovariance function $\langle x(t+\tau)\,x(t) \rangle \propto \tau^{-\beta}$, where $\beta=(\alpha-3)/2$ thus showing that we are dealing with a power-law correlated stochastic process. In the range $3<\alpha<5$ the process is long-range correlated \cite{Zoller}.

Let us now consider the coordinate transformation:
\begin{eqnarray}
          &&  x \mapsto y=f_l(x)= \left \{ \begin{array}{l}
                                \sqrt{2\,s}~{\rm{Erf}}^{^{-1}}
                                       \bigl[1-r(x) \bigl]     \hfill           ~|x| > L \\
                                                                                                   \\   
                                \sqrt{2\,s}~{\rm{Erf}}^{^{-1}}
                                       \bigl[1-p(x) \bigl]     \hfill           ~|x| \le L   
                       \end{array} \right.   \nonumber  \\
          &&    r(x)={
                      {2\,N_I\,x^{1-\alpha}}
                      \over
                      {\alpha-1}
                     }     \label{CHI_to_gauss}     \\
          &&    p(x)={{2\,N_I\,L} \over {L^\alpha} }-
                     {{N_I}\over{L^\alpha\,\cos(\sqrt{V_0} L)^2}}\,\bigl(x - L\bigr)+ \nonumber \\
          &&       -~{{N_I}\over{2 \sqrt{V_0}\,L^\alpha\,\cos(\sqrt{V_0} L)^2}} 
                     \Bigl(\sin(2 \sqrt{V_0} x)-  \sin(2 \sqrt{V_0} L)\Bigr)  \nonumber
\end{eqnarray}
By using the Ito lemma, one can show that, starting from the process of Eq. $(\ref{D1chimera})$, in the coordinate space $y=f_l(x)$ one gets a multiplicative stochastic process whose stationary pdf is given by Eq. $(\ref{pdfOU})$, with $s$ as an additional arbitrary parameter. One can show that $G(y)$ is continuos along the real axis, although its first derivative is discontinuos in $y=\pm f(L)$. The drift coefficient $H(y)$ is discontinuos in $y=\pm f(L)$. For large values of $y$ the drift and diffusion coefficient behave as:
\begin{eqnarray}
             && H(y) \propto {1 \over{y^{{\alpha+1}\over{\alpha-1}}}} \exp\bigl({-{y^2\over{s (\alpha-1)}}}\bigr) \qquad y \to +\infty, \nonumber \\
             && G(y) \propto {1 \over{y^{{\alpha}\over{\alpha-1}}}} \exp\bigl({-{y^2\over{2 s (\alpha-1)}}}\bigr) \qquad y \to +\infty. \nonumber
\end{eqnarray}

According to the eigenfunction expansion methodology \cite{risken,gardiner} the autocorrelation function $\rho_l(\tau)$ of the process defined by Eq. $(\ref{D1chimera})$ with the coordinate transformation of Eq. $(\ref{CHI_to_gauss})$ is given by Eq. $(\ref{Rrisk})$ with $f_s(x)$ replaced by $f_l(x)$. Now the functions $\psi_0(x)$ and $\psi_\lambda(x)$ are the eigenfunctions of the Schr\"odinger equation associated to the stochastic process of Eq. $(\ref{D1chimera})$ and with potential:
\begin{eqnarray}
              V_S(x)=\left \{\begin{array}{ccc}
                                   -V_0~    &~{\rm{if}}~&~|x| \leq L, \\
                                            &           &                  \\
                                   V_1/x^2 ~&~{\rm{if}}~&~|x| >   L,
                   \end{array} \right.  \label{VSchimera}
\end{eqnarray} 

\begin{figure} [t]
\begin{center}
              \resizebox{1\columnwidth}{!}{\includegraphics[scale=0.30] {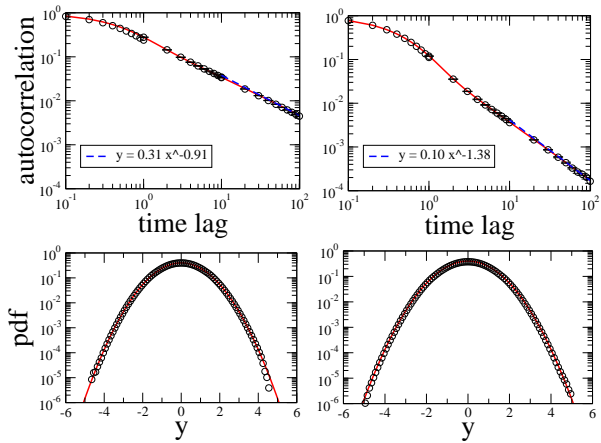} }
              \caption{(Color online) The figure shows time--average numerical simulations performed according to Eq. $(\ref{ACtimeens})$ for the case when when $s=1.0$, and $\alpha=3.05$ (left), $\alpha=4.00$ (right). In the top panel we show the results for the autocorrelation function. The solid (red) line shows the theoretical predictions obtained from Eq. $(\ref{Rrisk})$ with $f_s(x)$ replaced by $f_l(x)$, while the open circles show the result of the numerical simulations. The dashed (blue) lines show a nonlinear fit performed on the large time region of the autocorrelation function. The bottom panel shows the stationary pdf of the process. Again the solid (red) line shows the theoretical prediction of Eq. $(\ref{pdfOU})$, while the open circles show the result of the numerical simulations. The parameters of the numerical simulations are given in the text.}  \label{fig:CHIgauss}
\end{center}   
\end{figure}

In the top panels of Fig. $\ref{fig:CHIgauss}$ we show the results of numerical simulations of the autocorrelation function $\rho_l(\tau)$ performed for the case when $s=1.0$, $L=1.0$, and $V_0=0.98$ (i.e. $\alpha=3.05$, top-left panel) and $V_0=1.16$ (i.e. $\alpha=4.00$, top-right panel). The solid (red) lines show the theoretical prediction obtained from Eq. $(\ref{Rrisk})$ with $f_s(x)$ replaced by $f_l(x)$. The open circles show the results of the numerical simulations. By performing a nonlinear fit (dashed blue line), the autocorrelation function shows an asymptotic decay compatible with a power-law $\tau^{-\beta_l}$, with $\beta_l=0.91$ when $\alpha=3.05$ (top-left panel) and $\beta_l=1.36$ when $\alpha=4.00$ (top-right panel). The bottom panels of Fig. $(\ref{fig:CHIgauss})$ show the stationary pdf of the process. Again, the solid (red) line shows the theoretical prediction of Eq. $(\ref{pdfOU})$, while the open circles show the results of the numerical simulations for the case when $\alpha=3.05$ (bottom-left panel) and $\alpha=4.00$ (bottom-right panel).

By plotting the autocorrelation function $\rho_l(\tau)$ for $L=1.0$ and for different values of $V_0$, i.e. different values of $\alpha$, and by performing a nonlinear fit, one finds that for large lag values $\rho_l(\tau) \propto \tau^{-\beta_l}$ where $\beta_l$ shows a linear dependance from $\alpha$ described by $\beta_l=\alpha/2 - \eta$ with $\eta \approx 0.61$. By using the results of Ref. \cite{paperpa19} one obtains that indeed the autocorrelation function decays like $t^{-(\alpha-1)/2}$ with logarithmic corrections. When $\alpha<2\,(1+\eta)$ the process is long-range correlated.

Those shown in Fig. $\ref{fig:CHIgauss}$ are time--average numerical simulations performed according to Eq. $(\ref{ACtimeens})$. The parameters of the numerical simulations of the autocorrelation function are: $M=1$ in the region $\tau \leq 1$, $M=10$ in the region $\tau \in [1,10]$  and $M=20$ in the region $\tau \geq 10$ when $s=1.0$ and $\alpha=3.05$. The size of each time-series was: $T=10^{9}$ with a time-step of $\Delta t=0.005$ when $\tau<10$ and $T=10^{11}$ with a time-step was $\Delta t=0.005$ when $\tau>10$. When $s=1.0$ and $\alpha=4.00$ we used $M=1$ in the region $\tau \leq 1$, $M=20$ in the region $\tau >1$. The size of each time-series was: $T=10^{9}$ with a time-step of $\Delta t=0.005$ when $\tau<10$ and $T=10^{11}$ with a time-step was $\Delta t=0.005$ when $\tau>10$. The starting points of the simulated time-series were all the same with $x_j(0)=0.1$ where $j=1, \cdots, M$. The numerical simulations of the pdfs in the two bottom panels were obtained by using the same numerical data considered in the simulation of the autocorrelation function when $\tau>10$. In order to simulate the process in the $y$ coordinate space, we start by simulating the process of Eq. $(\ref{D1chimera})$ and compute $y=f_l(x)$ for each simulated $x$ value.

\subsection{The Doob Theorem}   \label{gaussdoob}

The existence of not exponentially correlated processes with Gaussian distribution does not contradict the Doob Theorem \cite{Doob}. In fact, such theorem deals with the case when the process admits a stationary pdf and a 2--point conditional transition probability that are both non singular and Gaussian on the whole real axis. Indeed the Doob theorem applies to the Ornstein-Uhlenbeck (OU) process described by the following linear Langevin equation \cite{risken}:
\begin{eqnarray}
          &&  \dot{x}(t)=-\gamma x(t)+ D \Gamma(t)   \label{D1OU}                 
\end{eqnarray}
and whose conditional probability is given by \cite{risken}: 
\begin{eqnarray}
          && \hspace{-0.6 cm}
                 P(x_2,t_2|x_1,t_1)={ 1 \over \sqrt{2 \pi D^2 (1-e^{- 2 \gamma \tau})}}~ \times \nonumber \\
          &&         \times~{\rm{exp}}\Bigl({ 1 \over{2 D^2 (1-e^{- 2 \gamma \tau})}}(x_2-e^{-\gamma \tau} x_1)^2
                                                                                                     \Bigr)    \label{GaussCondProb}
\end{eqnarray}
where $\tau=t_2-t_1$. The two parameters $D$ and $\gamma$ are related to the variance of the OU process by the relation $s=D^2/\gamma$.

A direct inspection of the conditional probabilities for the two processes of section \ref{riskgaussall} and \ref{chigaussall} shows that they are different from the one of Eq. $(\ref{GaussCondProb})$. Indeed, a preliminary issue regards the way to compare the three processes considered above. We first notice that, by construction, the three processes have the same stationary pdf of Eq. $(\ref{pdfOU})$. This is a Gaussian distribution with variance $s$. Hereafter we will always set $s=1$, which implies that $D=\sqrt{\gamma}$ in the OU process. As a result the OU process will be characterized by the single parameter $\gamma$, the process of section \ref{riskgaussall} will be characterized by the single parameter $k$ and the process of section \ref{chigaussall} will be characterized by the two parameters $L $ and $V_0$ or, by using Eq. $(\ref{pdfCHI})$, the two parameters $L$ and $\alpha$. Without loss of generality we can set $L=1$. Secondly, we notice that the three parameters $\gamma$, $k$ and $\alpha$ enter the autocorrelation function of the respective processes and therefore they determine how the processes loose or keep their memory. When comparing the three processes considered above one reasonable criterion for the choice of the parameters seems the impose that they carry the same amount of memory, i.e. the integral of the autocorrelation function is the same. Of course this is one out of many possible choices, and we are also aware of the fact that it can not be applied when considering long-range correlated processes. This choice however helps in presenting our results that are nevertheless based on analytical findings, as we will show in the next sections. 

\begin{figure} 
\begin{center}
              \resizebox{1\columnwidth}{!}{\includegraphics[scale=0.30] {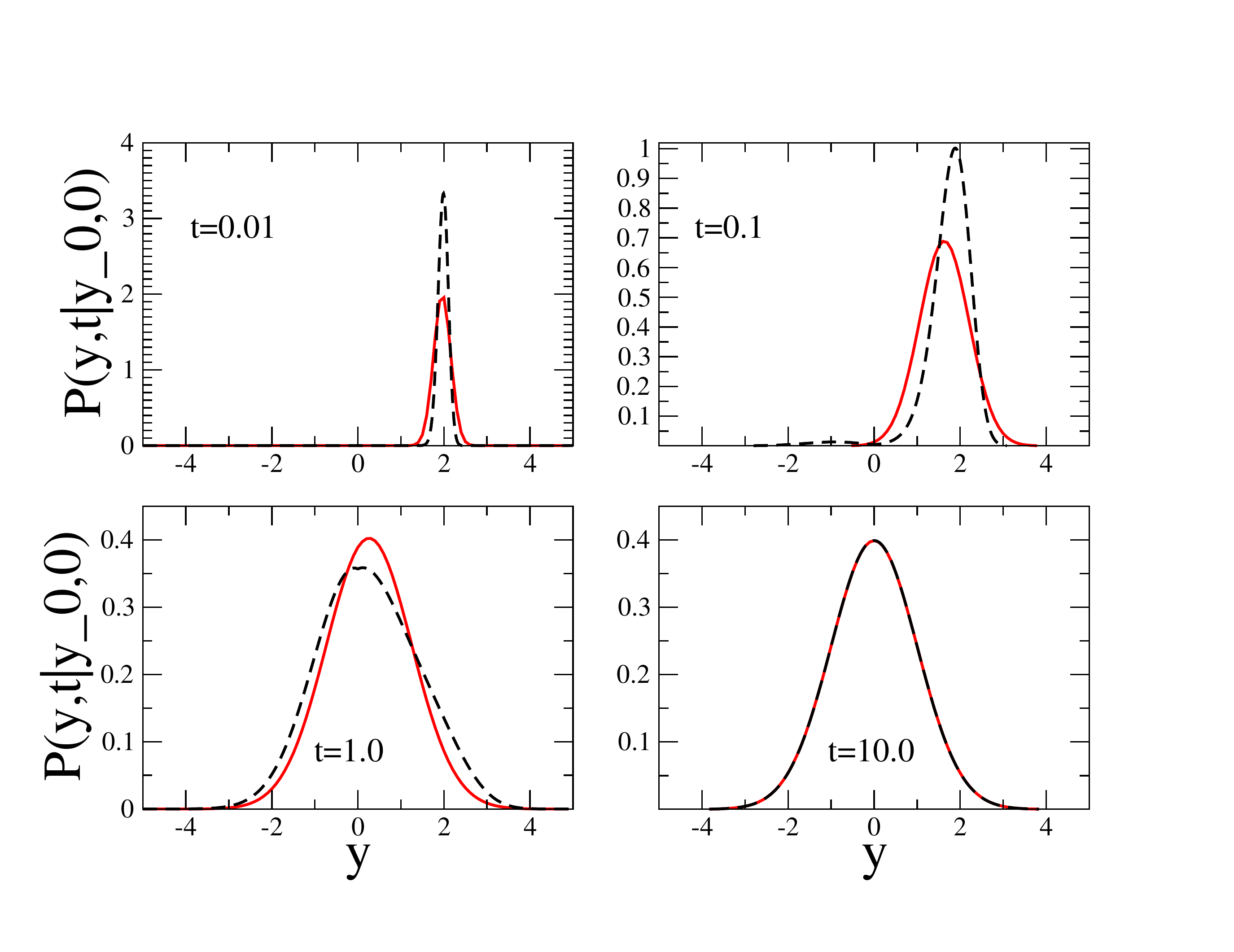} }
              \caption{(Color online) The figure shows a comparison between the conditional probability $P(y,t|y_0,0)$ of the stochastic process defined by Eq. $(\ref{D1risk})$ with the coordinate transformation of Eq. $(\ref{risken_to_gauss})$ and the one predicted for the OU process of Eq. $(\ref{GaussCondProb})$. Here, we consider the case when $s=1.0$, $k=2.0$ and $y_0=2.0$. The four panels show the behaviour of the conditional probability $P(y,t|y_0,0)$ with respect to $y$ for four different values of time $t$: $t=0.1$, $t=1.0$, $t=5.0$ and $t=10.0$. The open circles are the result of the numerical integrations of Eq. $(\ref{ProbCondEigen})$ and the solid (red) lines are the theoretical predictions obtained from Eq. $(\ref{GaussCondProb})$, where we used $\gamma=2.039$.}  \label{fig:2pTPrisk}
\end{center}   
\end{figure}

In order to illustrate how the existence of not exponentially correlated processes with Gaussian distribution does not contradict the Doob Theorem, in Fig. $\ref{fig:2pTPrisk}$ we show a comparison between the conditional probability $P(y,t|y_0,0)$ of the stochastic process defined by Eq. $(\ref{D1risk})$ with the coordinate transformation of Eq. $(\ref{risken_to_gauss})$ and the one predicted for the OU process. According to the above criterion, the parameters used for the process defined by Eq. $(\ref{D1risk})$ with the coordinate transformation of Eq. $(\ref{risken_to_gauss})$ are $s=1.0$, $k=2.0$ and $y_0=2.0$. For the OU process we considered $\gamma=2.039$. The four panels show the behaviour of the conditional probability $P(y,t|y_0,0)$ with respect to $y$ for four different values of time $t$: $t=0.1$, $t=1.0$, $t=5.0$ and $t=10.0$. The solid (red) lines are the theoretical predictions obtained from Eq. $(\ref{GaussCondProb})$ and the open circles are the result of the numerical integrations performed according to
\begin{eqnarray}
                   &&       P(y,t|y_0,0)=W_g(y)+ \label{ProbCondEigen}\\
                   &&       +\, {1 \over {\partial f_s(x)}}\,
                                 {{\psi_0(x)} \over {\psi_0(x_0)}}\,
                                 \int_{k^2/4}^{\infty} d\lambda \psi_\lambda(x)\,\psi_\lambda(x_0)\,e^{-\lambda t} \nonumber \\
                   &&       x=f_s^{-1}(y) \qquad \qquad x_0=f_s^{-1}(y_0) \nonumber
\end{eqnarray}
where $\psi_0(x)$ and $\psi_\lambda(x)$ are the eigenfunctions of the Schr\"odinger equation with potential $V_S(x)$$=k^2/4 - k\,\delta(x)$. Fig. $\ref{fig:2pTPrisk}$ shows that for small values of time, the conditional probability for the process of section \ref{riskgaussall} is quite different from the one of the OU process. As long as the time increases, the agreement gets better and better. In fact, it is expected that for large times both conditional probabilities must converge to the same pdf which is the one of Eq. $(\ref{pdfOU})$. The bottom-right panel of Fig. $\ref{fig:2pTPrisk}$ shows that at $t=10$ the agreement is already good.

In the next section we will analytically show that the way the conditional probability converges to the stationary pdf is indeed different for the three processes.

\section{Persistence}   \label{gausspersistence}

The existence of long-range interactions is often associated to the existence of persistencies in the time series of the stochastic process, i.e. to the fact that when a process reaches a position $x$ then it spends some time around that position value. The conditional probability gives the possibility to investigate the occurrence of persistencies. In fact, we consider the conditional probability at equal spatial position $P(x,t|x,0)$ as a proxy of the persistence of the process. 

\begin{figure} 
\begin{center}
              \resizebox{1\columnwidth}{!}{\includegraphics[scale=0.30] {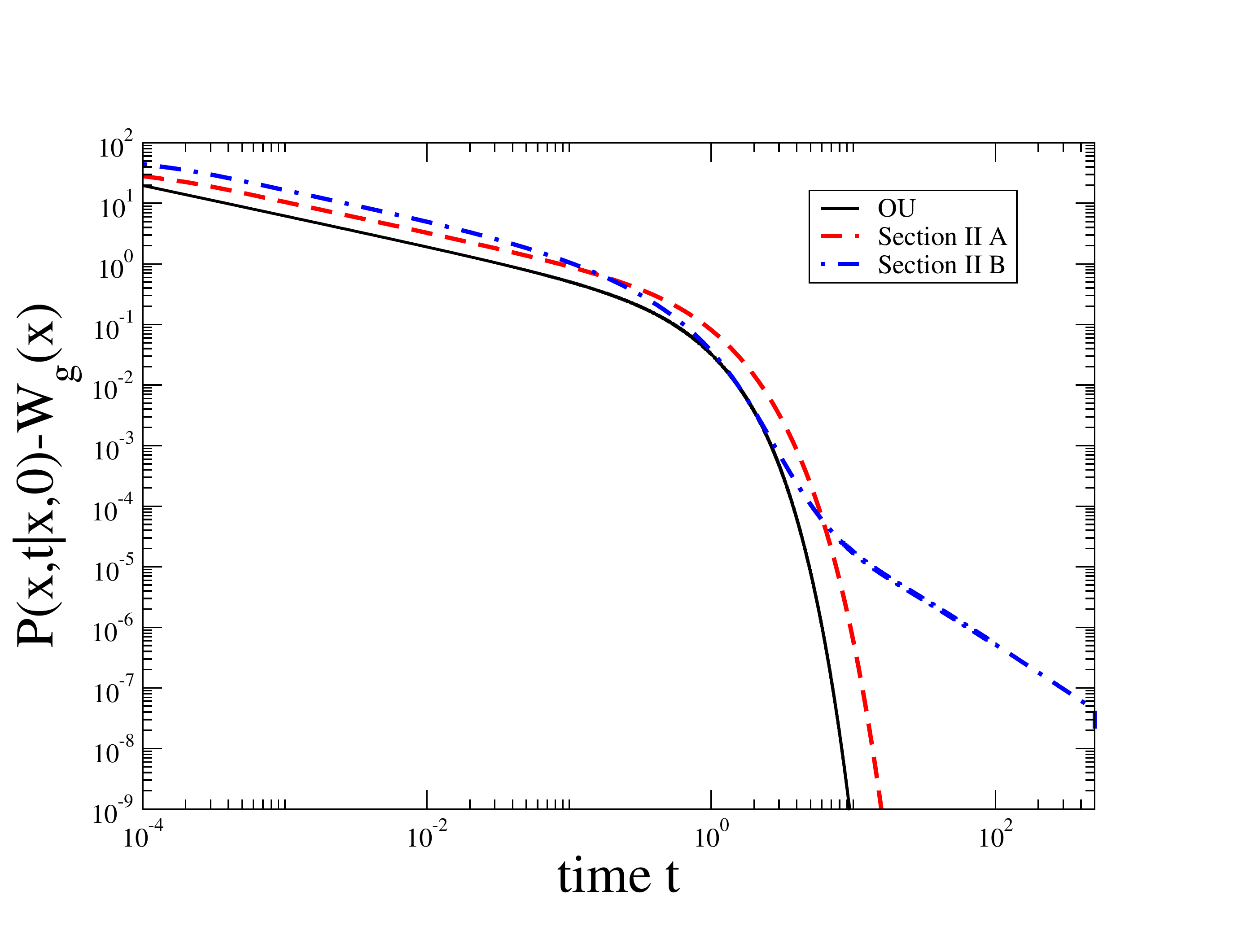} }
              \caption{(Color online) The figure shows a comparison between the subtracted conditional probability ${\cal{P}}_y(t)=P(y,t|y,0)-W_g(y)$ of $(i)$the OU process (black solid line) with parameters $\gamma=2.039$, $(ii)$ the process of section \ref{riskgaussall} (red dashed line) with parameters $k=2.0$, $s=1$ and $(iii)$ the process of section \ref{chigaussall} (blue dash-dotted line) with parameters $L=1.0$, $\alpha=4.00$, $s=1.0$. We show that whilst the OU process and the process of section  \ref{riskgaussall} shows a very fast decay, the ${\cal{P}}_y(t)$ relative to the process of section \ref{chigaussall} shows a slow decay that can be well described by a power-law.}  \label{fig:2pTPGauss}.
\end{center}   
\end{figure}

Starting from Eq. $(\ref{ProbCondEigen})$, we consider the quantity $P(y,t|y,0)$ for the processes of section \ref{riskgaussall} and section \ref{chigaussall}. In Fig. $\ref{fig:2pTPGauss}$ we show the subtracted quantity ${\cal{P}}_y(t) = P(y,t|y,0)-W_g(y)$ for $(i)$ the process of section \ref{riskgaussall} (dashed line) with $s=1.0$, $k=2.0$, and $(ii)$  section \ref{chigaussall} (dashed dotted line) with $s=1.0$, $L=1.0$, $V_0=1.183$, i.e. $\alpha=4.15$. The solid line shows the same quantity for the OU process and it is obtained from Eq. $(\ref{GaussCondProb})$ by simply imposing $x_2=x_1=y_0$. The parameters used here are $\gamma=2.039$. In all the three cases we considered $y=y_0=2.0$. The choice of these parameters ensures that the three processes have the same unitary variance and the same integral of the autocorrelation, i.e. they carry  the same ``amount of memory''. 

The figure shows how for small values of time the ${\cal{P}}_y(t)$ relative to the process of section \ref{chigaussall} is higher than the other two. The functional behaviour seems simular to the $1/\sqrt{t}$ behaviour which is typical of the OU case. In this range of time also the ${\cal{P}}_y(t)$ relative to the process of section \ref{riskgaussall} seems to share the same behaviour as the OU process. The regime that however is interesting for our purposes is the one of large times. Here we observe that whilst the OU process and the process of section  \ref{riskgaussall} shows a very fast decay, the ${\cal{P}}_y(t)$ relative to the process of section \ref{chigaussall} shows a very slow decay that can be well described by a power-law with ${\cal{P}}_y(t) \approx 1/t^{1.49}$. 

\begin{figure} 
\begin{center}
              \resizebox{1\columnwidth}{!}{\includegraphics[scale=0.30] {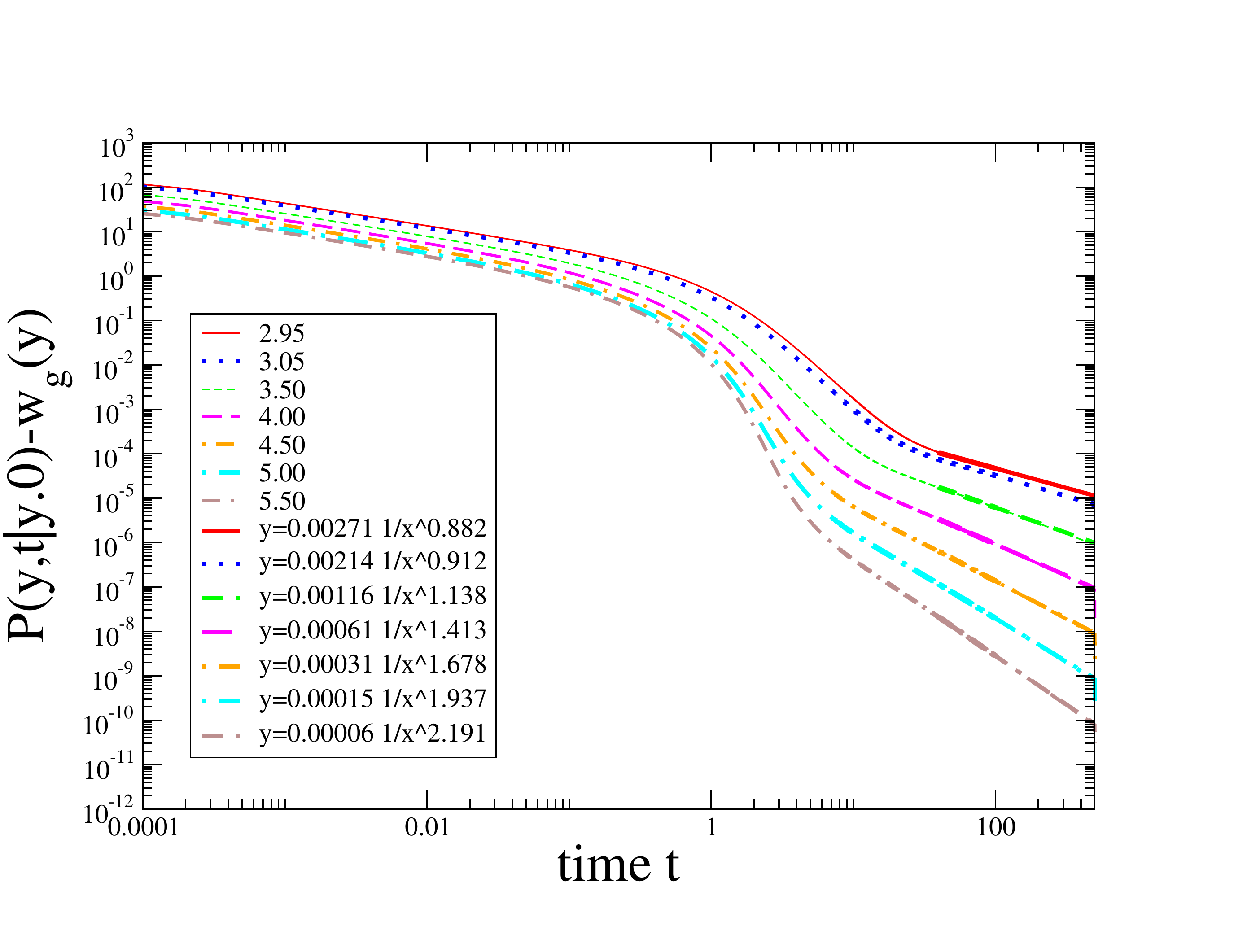} }
              \caption{(Color online) The figure shows the subtracted conditional probability ${\cal{P}}_y(t)=P(y,t|y,0)-W_g(y)$ for the process of section \ref{chigaussall} with parameters $L=1.0$, $s=1.0$ and different values of $\alpha$, as shown in the legend. In the bottom part of the legend we report the result of a nonlinear fit performed in the range $t \in[40,500]$. These results show that  ${\cal{P}}_y(t)$ shows an asymptotic power behaviour  ${\cal{P}}_y(t) \approx 1/t^\beta$, where $\beta$ seem to depend linearly from $\alpha$, according to the relation: $\beta=\alpha/2-\eta$. By using the results of asymptotic theory one can show that $\beta=(\alpha-1)/2$.}  \label{fig:2pTPChiGaussAll}.
\end{center}   
\end{figure}

In Fig. $\ref{fig:2pTPChiGaussAll}$ we show the function ${\cal{P}}_y(t)$ relative to the process of section \ref{chigaussall} for $s=1.0$, $L=1.0$ and different values of $V_0$ chosen  in such a way that the parameter $\alpha$ assumes the values shown in the legend. In the bottom part of the legend we report the result of a nonlinear fit performed in the range $t \in[40,500]$. These results show that  ${\cal{P}}_y(t)$ shows an asymptotic power behaviour  ${\cal{P}}_y(t) \approx 1/t^\beta$, where $\beta$ seem to depend linearly from $\alpha$, according to the relation: $\beta=\alpha/2-\eta$. 

Indeed, we are here interested in the large time behaviour of the subtracted persistence function ${\cal{P}}(t) \propto \int_0^\infty \psi_\lambda(x)^2 e^{- \lambda t}$, which is determined by the small $\lambda$ behaviour of $\psi_\lambda(x)^2$ \cite{Olver74}. One can show that in the limit when $\lambda \to 0$:
\begin{eqnarray}
             &&    \psi_\lambda(x) \approx     \lambda^{(\alpha-3)/4}  \label{SALcga}                
\end{eqnarray}
Therefore the large time behaviour of ${\cal{P}}(t)$ is given by:
\begin{eqnarray}
                {\cal{P}}_y(t) \approx  1/t^{(\alpha-1)/2}   \label{SALint}           
\end{eqnarray}
which is in good agreement with the results of Fig. $\ref{fig:2pTPChiGaussAll}$. Moreover, Eq. $(\ref{SALint})$ confirms that the conditional probability for such process is intrinsecally different from the one of Eq. $(\ref{ProbCondEigen})$ and therefore the Doob theorem can not be applied to this process.

Analogously, one can also show that in the case of the process of section \ref{riskgaussall}, one gets $\psi_\lambda(x) \approx   (\lambda- k^2/4)^{1/4} $ and therefore the subtracted persistence function ${\cal{P}}_y(t)$ shows a power-law decay with an exponential cut-off: ${\cal{P}}_y(t) \approx  e^{- {k^2 \over 4} t}\,t^{-3/2}$. Again, this result analytically confirms the ones of section \ref{gaussdoob}.

It is worth mentioning that the integral $\int_0^\infty dt \,{\cal{P}}_y(t)$ always exists for the OU process and the process of section \ref{riskgaussall}. For the process of section \ref{chigaussall} ${\cal{P}}_y(t)$ is integrable only if $\alpha  > 3$, i.e. when the pocess is short-range correlated. In this case, we can say that the conditional probability converges to the stationary distribution in a finite time. When $\alpha<3$, the process of section \ref{chigaussall} is long-range correlated, and also the above integral diverges, thus suggesting that there is not a typical time in which the conditional probability converges to the stationary distribution. Indeed, $\int_0^\infty dt \,{\cal{P}}_y(t)$ is related to the time spent by the process in a certain location. Such issue will be more quantitatively investigated in section \ref{gausspropagator}.

\subsection{Other processes with an infinite and unbounded set of timescales}   \label{coulgaussall}

The power-law correlated stochastic processes mentioned in the previous sections incorporates an infinite and unbounded set of timescales. This is due to the fact that the eigenvalues spectrum of the Schr\"odinger potential of Eq. $(\ref{VSchimera})$ is given by a continuum set of eigenvalues $\lambda>0$ attached to the null eigenvalue associated to the ground state. As mentioned above, the power-law correlated process shows a  power-law decay in the persistence function ${\cal{P}}_y(t)$, as illustrated in Fig. $\ref{fig:2pTPChiGaussAll}$. One might therefore wander whether or not such power-law decay is merely due to the fact that the process incorporates an unbounded set of timescales. In general, any Schr\"odinger potential that asymptotically decays like $1/x^\mu$ would give a stochastic process with an infinite and unbounded set of time-scales. In the case $\mu=1$, we will show here that the corresponding persistence function does not necessarily display a power-law decay, . 

Let us now consider the additive stochastic process associated to the Schr\"odinger potential \cite{risken}:
\begin{eqnarray}
              V_S(x)=\left \{\begin{array}{ccc}
                                   -V_0~    &~{\rm{if}}~&~|x| \leq L, \\
                                            &           &                  \\
                                   V_1/|x| ~&~{\rm{if}}~&~|x| >   L,
                   \end{array} \right.  \label{VScoulombwell}
\end{eqnarray} 
where $L$, $V_0$ and $V_1$ are real positive constants. This quantum potential is exactly solvable. In the region $x>L$ the eigenfunctions are given by:
\begin{eqnarray}
             && \psi_0(x)= A_0~\sqrt{x}~K_1(2\,\sqrt{V_1}~\sqrt{x}) \\
             && \psi_\lambda(x)= x~e^{- i \sqrt{\lambda} x}~\Bigl(                           
                                                                                    A_\lambda\,F_{1,1}(1 - i {V_1 \over 2}~{1 \over \sqrt{\lambda}}, 2, 2 i  \sqrt{\lambda} x)+ \nonumber \\
             &&         \hspace{2.5 truecm}                    C_\lambda\,U(1 - i {V_1 \over 2}~{1 \over  \sqrt{\lambda}}, 2, 2 i  \sqrt{\lambda} x)
                                                                           \Bigl)+c.c.         \nonumber 
\end{eqnarray}
where $K_1(\cdot)$ is the K-Bessel function of order $1$, $U(\cdot)$ is the confluent hypergeometric function and $F_{1,1}(\cdot)$ is the Kummer confluent hypergeometric function. Usually the $|x|^{-1}$ potentials arise in the context of quantum systems subject to a repulsive or attractive central Coulomb potential \cite{Landau}. The Coulomb potential is then defined only for positive real values and the relevant eigenfunction must be regular at the origin. In our case, we are considering a Coulomb potential extending over the whole real line although it is regularized in $x=0$ by assuming the $V_S(x)=-V_0$ when $|x|<L$.  This implies that the contribution associated to the confluent hypergeometric function $U(\cdot)$ must be taken into account. 

The normalization constant $A_0$ is fixed by imposing that the ground state is normalized to unity. The other normalization constants $A_\lambda$ and $C_\lambda$ are fixed by imposing that the odd and even eigenfunctions fulfill the normalization condition $\int dx\, \psi_\lambda(x) \psi_{\lambda'}(x)=\delta(\lambda-\lambda')$. The drift coefficient $h(x)$ aymptotically decays like $h(x) \approx 1/\sqrt{x}$.

As mentioned above, the large time behaviour of the subtracted persistence function ${\cal{P}}(t)$  is determined by the small energy behaviour of $\psi_\lambda(x)^2$ \cite{Olver74}. One can show that in the limit when $\lambda \to 0$:
\begin{eqnarray}
                \psi_\lambda(x) \approx  e^{- { {3 \pi V_1} \over {4 \sqrt{\lambda}}}}   \label{psismallCW}           
\end{eqnarray}
and therefore the large time behaviour of ${\cal{P}}_x(t)$ is:
\begin{eqnarray}
               {\cal{P}}_x(t)     \approx  {{e^{- \kappa t^{1/3}}} \over {t^{5/6}}} \qquad \kappa=3\,\Bigl( { { 3\pi \over 4} V_1} \Bigl)^{2/3}
\end{eqnarray}
The above result shows that the mere existence of an infinite and unbounded set of timescales is not enough to ensure an asymptotic power-law behaviour in ${\cal{P}}_x(t)$. It is worth mentioning that such result is also valid for any coordinate transformation able to map the stochastic process associated to the quantum potential of Eq. $(\ref{VScoulombwell})$ into a Gaussian distributed one. In passing, it is also possible to prove that the stochastic process associated to the quantum potential of Eq. $(\ref{VScoulombwell})$ is not power-law correlated, even though it incorporates an infinite and unbounded set of timescales \cite{farago}. 

\section{Mean First Passage time}   \label{gausspropagator}

Na\"ively speaking, the existence of power-law tails in ${\cal{P}}_y(t)$ indicates that the process with power-law autocorrelation function spends more time than the other two processes in a certain spatial position. This would imply that, starting from position $x$ the power-law autocorrelated process takes more time than the others to reach another position at distance $\Lambda$ from $x$. In other words the process with power-law autocorrelation function is slower than the other two processes. Such an effect can be quantitatively measured by considering the mean First Passage Time \cite{FPT1,FPT2,MET}.

Specifically, we will consider the mean time $T_{x}(\Lambda)$ that is needed to reach for the first time position $x \pm \Lambda$ starting from position $x$.  This is obtained by solving the equation \cite{gardiner}:
\begin{eqnarray}
                             h(x) { \partial T_{x}(\Lambda) \over \partial x} + g(x)^2 { \partial^2 T_{x}(\Lambda) \over \partial x^2} = -1 \label{FPTequation}
\end{eqnarray}
with boundary conditions $T_{x}(\Lambda)=0$ when $x=\pm \Lambda$. Here $h(x)$ and $g(x)$ are the drift and diffusion coefficient appearing in the non linear Langevin equation associated to the considered stochastic process.

In the case of the OU process of Eq. $(\ref{D1OU})$ the mean First Passage Time is given by:
\begin{eqnarray}
                               T_0(\Lambda)={\Lambda ^2 \over {2 D^2}} \,{\rm{H_{PFQ}}}\Bigl( \{1,1\},\{3/2,2\}, \gamma {\Lambda ^2 \over {2 D^2} }\Bigl) \label{FPTou}
\end{eqnarray}
with $D=\sqrt{\gamma\,s}$.

For the stochastic Gaussian process of section \ref{riskgaussall} the mean First Passage Time can be analytically computed by using the drift coefficient of Eq. $(\ref{hRGA})$ and the diffusion coefficient of Eq. $(\ref{gRGA})$ into Eq. $(\ref{FPTequation})$. For the case when $x=0$ one gets:
\begin{eqnarray}
                      &&       T_0(\Lambda)= {1 \over k^2} \,\log\Bigl(1- {\rm{Erf}}\bigl(\Lambda/\sqrt{2 s}\bigl) \Bigl)\,+\label{FPTRGA} \\
                      &&        \hspace{2.5 truecm} -\,{1 \over k^2}\Bigl(1- {1 \over{1- {\rm{Erf}}\bigl(\Lambda/\sqrt{2 s}\bigl)}}\Bigl) \nonumber
\end{eqnarray}
An alternative way to obtain such result is to analytically compute the mean First Passage Time for the stochastic process of Eq. $(\ref{D1risk})$ and then performing the substitution:
\begin{eqnarray}
                      \Lambda \mapsto - {1 \over k} \log\Bigl(1- {\rm{Erf}}\bigl(\Lambda/\sqrt{2 s}\bigl) \Bigl)
\end{eqnarray}
For large values of $\Lambda$ the mean First Passage Time of Eq. $(\ref{FPTRGA})$ gives:
\begin{eqnarray}
                      &&       T_0(\Lambda) \approx z \, e^{z^2} \qquad z={\Lambda \over \sqrt{2 s}} \label{FPTRGAas}
\end{eqnarray}

For the stochastic Gaussian process of section \ref{chigaussall} the mean First Passage Time can not be computed by analytically solving Eq. $(\ref{FPTequation})$. However, one can follow the alternative way mentioned above. Specifically, we first computed analytically the mean First Passage Time of the process of Eq. $(\ref{D1chimera})$ by using Eq. $(\ref{FPTequation})$. Subsequently, according to Eq. $(\ref{CHI_to_gauss})$ we performed the substitution:
\begin{eqnarray}
                      &&    \Lambda \mapsto f_I^{-1}(\Lambda)        \qquad \qquad {\rm{if}}  \,  \Lambda>L \\
                      &&    \Lambda \mapsto f_{II}^{-1}(\Lambda)     \qquad \qquad {\rm{if}}  \,  \Lambda<L
\end{eqnarray}
In the region $\Lambda>f_I(L)$ it is possible to obtain an analytical expression showing that for large values of $\Lambda$ one gets:
\begin{eqnarray}
                      &&       T_0(\Lambda) \approx \Bigl(  z \, e^{z^2} \Bigl)^{{\alpha+1}\over{\alpha-1}} \qquad z={\Lambda \over \sqrt{2 s}} \label{FPTCGAas} 
\end{eqnarray}
Eq. $(\ref{FPTCGAas})$ and Eq. $(\ref{FPTRGAas})$ clearly show that for large values of $\Lambda$ the process of section \ref{chigaussall} has a mean First Passage Time larger than the process of section \ref{riskgaussall}. This confirms that the process that incorporates more time-scales is the slowest.

For small values of $\Lambda $ in all considered processes the mean First Passage Time grows quadratically. In fact, from Eq. $(\ref{FPTou})$ one easily gets that the OU process admits $T_0(\Lambda) \approx {\Lambda^2 /{(2 \gamma s)}}$ while for the process of section \ref{riskgaussall} from Eq. $(\ref{FPTRGA})$ one easily gets $T_0(\Lambda) \approx \Lambda^2/(\pi s k^2)$ in the limit when $\Lambda \to 0$. For the process of section \ref{chigaussall} we do not have an analytical expression for $T_0(\Lambda)$, since $y=f_{II}(x)$ is not analytically invertable although it is continuous and monotonic and therefore its inverse function does exist. However, one can show that for small values of $x$ one gets $y=f_{II}(x) \approx K x$. where:
\begin{eqnarray}
                               K={{4\,\sqrt{2 \pi s}\,(\alpha-1) L V_0} \over {4 L^2 V_0- \alpha (1 - 4 L^2 V_0) - \alpha\,\cos(2 L \sqrt{V_0})}}
\end{eqnarray}
By using the mean First Passage Time of the process of Eq. $(\ref{D1chimera})$ and by performing the substitution $\Lambda \mapsto \Lambda/K$ for small values of $\Lambda$ one gets $T_0(\Lambda)\approx\Lambda ^2/(2 K^2)$.

\begin{figure} 
\begin{center}
              \resizebox{1\columnwidth}{!}{\includegraphics[scale=0.30] {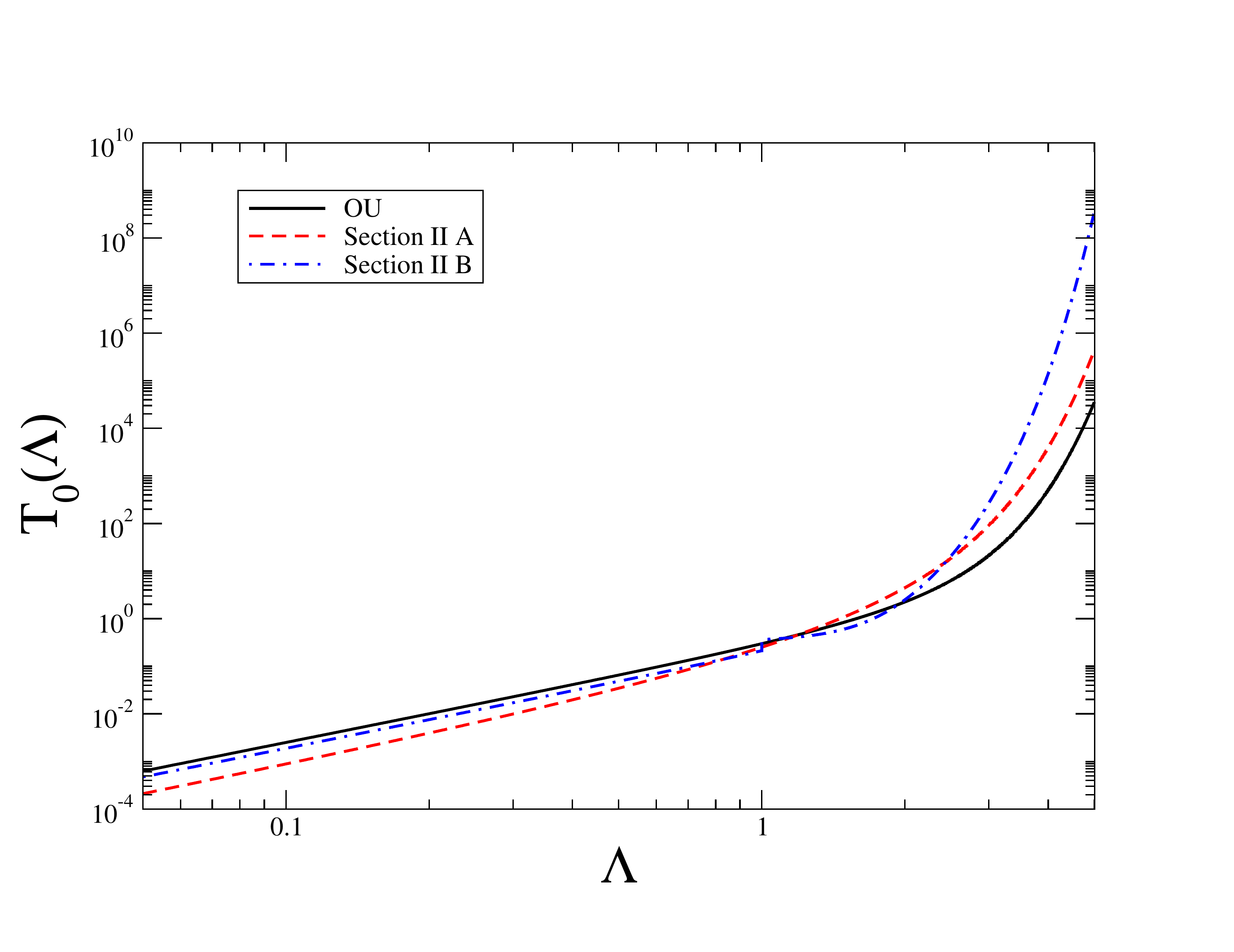} }
              \caption{(Color online) The figure shows a comparison between the mean First Passage Time $T_x(L)$ of $(i)$the OU process (black solid line) with parameters $\gamma=2.039$, $(ii)$ the process of section \ref{riskgaussall} (red dashed line) with parameters $k=2.0$, $s=1$ and $(iii)$ the process of section \ref{chigaussall} (blue dash-dotted line) with parameters $L=1.0$, $\alpha=4.00$, $s=1.0$. We show that for large values of $\Lambda$ the more time-scales the process includes, the larger the mean First Passage Time. With this choice of the relevant parameters, although both processes of section \ref{riskgaussall} and section \ref{chigaussall} are more persistent than the OU one, for small values of $\Lambda$ they can be faster than OU.}  \label{fig:FPTGauss}.
\end{center}   
\end{figure}

In Fig. $\ref{fig:FPTGauss}$ we show the mean First Passage Time for the two processes of section \ref{riskgaussall} (dashed line), with parameters $s=1.0$, $k=2.00$, and section  \ref{chigaussall} (dashed dotted line), with parameters $s=1.0$, $L=1.0$, $V_0=1.16$, i.e. $\alpha=4.00$. For comparison, the solid line shows the same quantity for the OU process, with $D=\sqrt{\gamma}$, $\gamma=2.039$. Again, the choice of these parameters ensures that the three processes have the same unitary variance and the same integral of the autocorrelation, i.e. they carry  the same ``amount of memory''.  

As expected, for large values of $\Lambda$ the more time-scales the process includes, the larger the mean First Passage Time. Surprisingly, with this choice of the relevant parameters, although both processes of section \ref{riskgaussall} and section \ref{chigaussall} are more persistent than the OU one, for small values of $\Lambda$ they can be faster than OU. This suggests that when a process statistically spends more time in a given position, say $x=0$, the likelihood that it reaches nearby positions by chance is also enhanced. This is another way of looking at persistence.

\section{Conclusions}   \label{concl}

The aim of the paper was to  understand how the inclusion of more and more time-scales into a stochastic process affects its conditional probability. We recall that, together with the stationary distribution, this is the other probability distribution that fully characterizes a stationary Markovian process. 

In order to disentangle all possible effects due to the stationary distribution, we have considered processes with a Gaussian distribution and characterized by the presence of $(i)$ one single time-scale, i.e. the OU process, $(ii)$ an infinite set of time-scales bounded from below that give rise to a short-range correlated process, see section \ref{riskgaussall}, and $(iii)$ an infinite and unbounded set of time-scales that give rise to a power-law correlated process, see section \ref{chigaussall}.

For these processes we have considered the equal position conditional probability $P(x,t|x,0)$ and the mean First Passage Time $T_x(\Lambda)$. The function $P(x,t|x,0)$, or better ${\cal{P}}_x(t)=P(x,t|x,0)-W_g(x)$, tells us how the conditional probability converges to the stationary Gaussian distribution $W_g(x)$. Moreover, $P(x,t|x,0)$ can be considered as a proxy of the persistence, i.e. the fact that when a process reaches a position $x$ then it spends some time around that position value. The Mean First Passage Time  tells us how much time needs a process to reach a position at distance $\Lambda$ starting from position $x$.

In section \ref{gausspersistence} we have shown that the more time-scales the process includes, the larger the persistence. Specifically, we have shown that the power-law correlated process of section \ref{chigaussall} shows a slow power-law decay of ${\cal{P}}_y(t)$. By contrast, the other two processes show a convergence dominated by an exponential cut-off. Moreover, in section \ref{coulgaussall} we have shown that the existence of an infinite and unbounded set of timescales is a necessary but not sufficient condition for observing the power-law decay of ${\cal{P}}_y(t)$.

We have therefore given evidence that the existence of persistencies, as measured by ${\cal{P}}_y(t)$, are neither necessarily related to existence of extreme events \cite{Embrechts, Havlin}, nor to the existence of an infinite set of time-scales in the process. In the context of stationary Markovian processes the persistence seems to be associated to the power-law decay of the autocorrelation function, as for the process of section \ref{chigaussall}. This means that the weights by which each time-scale enters the process play a fundamental role. Moreover, it is worth mentioning that the power-law decay of ${\cal{P}}_y(t)$ can be observed for all values of $\alpha$, i.e. when the process is either short-range and long-range power-law correlated. However, when the process is long-range correlated, the integral $\int_0^\infty dt \, {\cal{P}}_y(t)$ diverges.

In section \ref{gausspropagator} we have shown that for large values of $\Lambda$ the more time-scales the process includes, the larger the mean First Passage Time. Specifically, we have shown that the process of section \ref{chigaussall} shows a power-law growth while the process of secton \ref{riskgaussall} shows a linear growth in terms of the $\bigl(z \, e^{z^2}\bigl)$ function, with $z=\Lambda / \sqrt{2 s}$. This indicates that, starting from position $x=0$, locations that are very distant are reached with increasing difficulty. That raises the problem of the effective capacity of a power-law correlated process to span all the real axis in a finite time. Surprisingly,  for small values of $\Lambda$, at least for some choices of the relevant parameters, processes that include more time-scales show a smaller mean First Passage Time. This suggests that when a process statistically spends more time in a given position, say $x=0$, the likelihood that it reaches nearby positions by chance is also enhanced.

Further work is needed to understand how the above results are related to the ergodicity properties \cite{Lutz} of the process of section \ref{chigaussall}.  Specifically, a issue worth of further investigation is how $(i)$ the divergence of $\int_0^\infty dt \,{\cal{P}}_y(t)$ when $\alpha<3$ and $(ii)$ the power-law growth of the mean First Passage Time affect the fact that the process is ergodic or not.

\section*{Aknowlegments}
We are grateful to Dr. E. d'Emilio for the enlightening discussions about the analytical solutions of the quantum potentials of section \ref{chigaussall} and section \ref{coulgaussall}.


\end{document}